\documentclass{article}
\usepackage[T1]{fontenc}
\usepackage[latin1]{inputenc}
\usepackage{a4wide}

\makeatletter

\newcommand{\LyX}{L\kern-.1667em\lower.25em\hbox{Y}\kern-.125emX\spacefactor1000}

\makeatother

\begin{document}

\title{Information - The Fundamental Notion of Quantum Theory}

\author{Gesche Pospiech}

\maketitle
\begin{abstract}
In this article a notion of information is presented which stresses the contextuality
of quantum objects and their measurement. Mathematically this is reached by
a quantification of the quantum mechanical surplus knowledge which has been
introduced by Weizsäcker. This new formulation gives insight into the relation
between single quantum objects and ensembles of quantum objects. The goal is
to provide an explanatory concept for teaching purposes, the description of
quantum processes and measurement with aid of information.
\end{abstract}

\section{Introduction}

Recently, the experiments performed by the groups of Zeilinger and Weinfurter
during the last years indicate that in understanding quantum theory the concept
of information may play an essential role. Zeilinger proposed in \cite{zeil_inform}
an information measure suitable for use in quantum physics.

Here I try to introduce an information measure taking into account two main
points of the discussion about interpretation of quantum theory:

\begin{enumerate}
\item The contextuality of any quantum measurement.
\item The transition from quantum regime to classical regime taking place during a
measurement.
\end{enumerate}
Because of the significance of the first point I will concentrate on it. Already
Niels Bohr stressed in his interpretation the importance of taking into account
the interdependance of measuring apparatus and quantum object. He always stated
that in every measurement also the apparatus has to be described exactly in
order to account for the observable properties of the quantum object in question.
Only the observers' questions decide which of the quantum objects' property
gets a precise well-defined value. Other properties related to it by an uncertainty
relation, however, are still indetermined

\section{Some basics}

In this section I shortly describe some features of the density matrix formalism
as far as necessary for my purpose. I restrict to the case of finite dimensional
Hilbert space.

Two different types of quantum objects can be distinguished, which I classify
according to the goal, the description of measurement in terms of information.
The key notion is the description of quantum objects by a density matrix.

\subsection{Quantum Objects in Pure State}

I denote a single quantum object, unknown to the environment by the term \emph{type
1-system}; such an object has never been measured. It likewise could be described
by a \( \psi  \)-function, written as \( \psi =\sum ^{n}_{i=1}a_{i}\varphi _{i} \)
, \{ \( \varphi _{i} \)\} an orthonormal basis of the underlying Hilbert space.
Then its density matrix \( \widehat{\rho } \) has the form:\\

\begin{equation}
\label{dichte1}
\rho =\left( \begin{array}{ccccc}
\left| a_{1}\right| ^{2} & a_{1}a_{2}^{*} & a_{1}a_{3}^{*} & ... & ...\\
a_{2}a_{1}^{*} & \left| a_{2}\right| ^{2} & a_{2}a_{3}^{*} & ... & ...\\
a_{3}a_{1}^{*} & a_{3}a_{2}^{*} & \left| a_{3}\right| ^{2} & ... & ...\\
... & ... & ... & ... & ...\\
... & ... & ... & ... & ...
\end{array}\right) \quad \mbox {with}\quad tr\rho =\sum _{i=1}^{N}\left| a_{i}\right| ^{2}=1
\end{equation}
 Such a hermitian matrix can be brought to diagonal shape by an unitary transformation
to e.g.:
\begin{equation}
\label{dichte_qcl}
\widehat{\rho ^{T}}=\left( \begin{array}{ccc}
1 & 0 & ..\\
0 & 0 & ..\\
.. & .. & ..
\end{array}\right) 
\end{equation}

This transformation could in principle be written as \( U=e^{-\imath Ht} \)
with a suitable Hamiltonoperator \( H \) and a suitable time \( t \). But
since the original density matrix is unknown, also the transformation can not
be known in advance. There only could be an educated guess in order to achieve
this form of the density matrix \label{sec: unitary}. This technique is used
in the development of quantum computing (Grovers algorithm)(see e.g.\cite{exp_qsearch})
.

\subsubsection{Contextuality of quantum objects\label{sec: contextuality}}

The above matrix \( \rho  \) represents a pure state, i.e. \( \rho ^{2}=\rho  \).
This property is invariant under unitary transformations. But the coefficients
in the matrix representation depend on the basis chosen. We imagine that the
state \( \psi  \) resp. \( \rho  \) is represented with respect to a specified
measuring apparatus which means selection of a measurement basis \( \left\{ \varphi _{i}\right\} . \)
Hence the entries of the matrix reflect the relation between the state and the
chosen measurement, or in other words, the quantum object in its context. If
the density matrix looks like \( \widehat{\rho ^{T}} \) (see (\ref{dichte_qcl})),
then the state is an eigenstate with respect to the measurement basis. This
can be interpreted as the quantum object being in a definite state relative
to the corresponding measurement, i.e. the corresponding eigenvalue is attained
with probability 1. If this is not the case the density matrix will be of shape
(\ref{dichte1}) with at least two of the \( a_{i}\neq 0 \).

\paragraph{Example: }

The simplest possible example is a single photon with density matrix 
\[
\rho _{z}=\left( \begin{array}{cc}
a_{1}a_{1}^{*} & a_{1}a_{2}^{*}\\
a_{1}^{*}a_{2} & a_{2}a_{2}^{*}
\end{array}\right) \]
 relative to the basis formed by the eigenstates of the \( \sigma _{z} \)-Operator,
say. It would look different with respect to the eigenstates of the \( \sigma _{x} \)-Operator,
namely: 
\[
\rho _{x}=\frac{1}{2}\left( \begin{array}{cc}
(a_{1}+a_{2})(a_{1}+a_{2})^{*} & (a_{1}+a_{2})(a_{1}^{*}-a_{2}^{*})\\
(a_{1}+a_{2})^{*}(a_{1}-a_{2}) & (a_{1}-a_{2})(a_{1}-a_{2})^{*}
\end{array}\right) \]
\label{sec: beispiel}

\subsection{Ensemble of Quantum Objects }

With the term \emph{type 2-system} I denote an ensemble of quantum objects.
The density matrix of such an ensemble is described by diagonal entries giving
the probabilities of the corresponding measurement results (\cite{v_hund})
and hence might be written as 
\begin{equation}
\label{dichte_ensemble}
\widetilde{\rho }=\left( \begin{array}{ccccc}
\left| a_{1}\right| ^{2} & 0 & 0 & 0 & ...\\
0 & \left| a_{2}\right| ^{2} & 0 & 0 & ...\\
0 & 0 & \left| a_{3}\right| ^{2} & 0 & ...\\
... & ... & ... & ... & ...\\
... & ... & ... & ... & ...
\end{array}\right) \quad \mbox {with}\quad tr\rho =\sum _{i=1}^{N}\left| a_{i}\right| ^{2}=1
\end{equation}
 In this case the behaviour of the quantum objects allows for an ignorance interpretation,
where the properties of a single object are fixed but unknown. Therefore it
could be modelled by throwing dice.

These two cases can be distinguished by a proper notion of quantum information.

\section{Notion of Quantum Information}

\subsection{Quantum Information}

\begin{description}
\item [Definition:]The quantum information, i.e. the information present in a quantum
object, will be defined as 
\begin{equation}
\label{def_inf}
I_{Q}:=C\, tr\rho ^{2}\quad C\, \mbox {a\, suitably\, chosen\, constant}
\end{equation}

\end{description}
This defintion is motivated by the fact, that the density matrix incorporates
all the properties of the object in question. The expectation value of any observable
\( O \) is given by: \( \left\langle O\right\rangle :=tr(\rho O) \). Hence
the expression \( tr \)\( \rho ^{2} \) can be considered as the expectation
value of the information inherent in the quantum object. The normalization constant
\( C \) should be chosen as \( \log N \), where \( \log  \) denotes the logarithm
of basis 2 and \( N \) is the dimension of the underlying Hilbert space. In
this formulation the value \( C \) corresponds to the whole available information,
counted in bit or put in other words, the minimal number of questions necessary
for determining the state of the quantum object. The value 0 - that cannot be
attained - would correspond to the (impossible) case that there is no (quantum)
information at all.

\subsubsection{Type 1 Systems}

From the definition it is clear that \( I_{Q}=C=\log N \) for every quantum
object in a pure state (since \( \rho ^{2}=\rho  \)). It is interesting to
give a meaning to this result: A quantum object is, in strength, completely
isolated from its environment. Hence this result can be interpreted in a way
that the quantum object has the whole information - including its quantum mechanical
surplus-knowledge (see below, \ref{sec: spk}) - incorporated on its own. And
this internal information of the quantum object is independent of all other
things that might happen in the world. Even changes internal to the quantum
object do not have any influence on the amount of information as long as they
correspond to unitary transformations. The quantum information is - by the very
definition - always equal to \( C \), i.e. always complete and always the whole
information thinkable of.

\subsubsection{Type 2-systems}

In the case of type 2-systems, however, \( I_{Q}<C \) in general. Here the
definition of quantum information gives - with respect to a suitable basis
\[
I_{Q}=Ctr\widetilde{\rho }^{2}=C\sum _{i}\left| a_{i}\right| ^{4}>0\]
 where \( \left| a_{i}\right| ^{2} \)are the diagonal elements of the density
matrix \( \widetilde{\rho } \) (see (\ref{dichte_ensemble})). This measure
of information attains a minimum \( \frac{\log N}{N} \) if all states are equally
probable, i.e. \( \left| a_{i}\right| ^{2}=\frac{1}{N} \) for all \( i \),
and a maximum, namely \( \log N \), if one state is attained with probability
1. 

The difference between the two types of quantum systems hence is clearly visible
on the basis of the notion of quantum information. This instrument can be more
refined.

\subsection{Interaction of quantum objects\label{sec: interaction}}

Let two quantum objects - the object S and the object M with density matrices
\( \rho _{S} \); \( \rho _{M} \) respectively - interact with each other.
In this interaction case the definition
\[
I^{I}_{Q}:=(C_{M}+C_{S})tr(\rho _{S}\otimes \rho _{M})^{2}\]
 gives the information, object M carries about object S, say. \( I_{Q}^{I} \)
, by definition, does not change during the interaction as long as it is described
by a Schrödinger equation.

We can distinguish three cases:

\begin{enumerate}
\item Both objects, S and M, are of type 1. Then the compound system is again of type
1, and an isolated quantum object with quantum information 
\[
I^{I}_{Q}=(C_{M}+C_{S})tr(\rho _{S}\otimes \rho _{M})^{2}=C_{M}+C_{S}\]
This means, in the context of two quantum objects, that both carry the full
information of \emph{each} \emph{other} because of the entanglements arising
between them because of the interaction. 
\item Both objects, S and M, are of type 2. Then the compound system is again of type
2 with quantum information 
\[
I^{I}_{Q}=(C_{M}+C_{S})tr(\rho _{S}\otimes \rho _{M})^{2}=(C_{M}+C_{S})\sum _{i}\left| a_{i}^{S}\right| ^{4}\sum _{j}\left| a_{j}^{M}\right| ^{4}=\frac{(C_{M}+C_{S})}{C_{S}C_{M}}I_{Q}^{S}I_{Q}^{M}\]
 where \( I_{Q}^{S};I_{Q}^{M} \) are the quantum information of object S and
object M, respectively. 
\item The third case is the most interesting case because it can be used for a characterization
of measurement: The object S is of type 1 and the object M of type 2. We have:
\[
I^{I}_{Q}=(C_{M}+C_{S})tr(\rho _{S}\otimes \rho _{M})^{2}=(C_{M}+C_{S})\frac{I^{S}_{Q}}{C_{S}}\cdot \frac{I^{M}_{Q}}{C_{M}}=I_{Q}^{M}(1+\frac{C_{S}}{C_{M}})<C_{M}+C_{S}\]
 The strict inequality indicates that the measuring object M takes information
from object S, but in general not the whole information \( I_{Q}^{S}=C_{S} \).
(Of course M still holds its ``own'' information \( I_{Q}^{M} \).)
\end{enumerate}

\subsection{The quantum mechanical surplus-knowledge\label{sec: spk}}

The information \( I_{Q} \) only ``sees'' the quantum object, not any relation
to a measurement. Its constant value \( C \) for a pure state reflects the
fact that a quantum object always carries the whole information about its state
in it. In a measurement, however, only parts of this information come into ``reality''.
The other parts are called the ``quantum mechanical surplus-knowledge'' by
Weizsäcker, \cite{aufbau}, and Görnitz \cite{goernitz} stresses the importance
of the relations between different parts of a quantum object. Hence the off-diagonal
elements of the density matrix seem to be an appropriate measure for this ``surplus
knowledge''. As alluded to before, (see section \ref{sec: unitary}), the off-diagonal
elements depend on the kind of contact with environment (measurement) or, in
other words, on the relation between the state of the quantum object and the
(planned) measurement. This observation also reflects the considerations of
Bohr who always stressed that the appearance of a quantum object depends on
the kind of measurement. The surplus-knowledge hence is deeply connected to
the basis chosen, i.e. to the ``planned measurement''. 

How to define the surplus-knowledge? Let \( \rho  \) be the density matrix
of a quantum object S of shape (\ref{dichte1}) and \( \widetilde{\rho } \)
the corresponding diagonal matrix \( \widetilde{\rho }=diag_{i}(\left| a_{i}\right| ^{2}) \). 

The relation between \( \rho  \) and \( \widetilde{\rho } \) can be interpreted
in a twofold way: 

\begin{enumerate}
\item Given a  quantum object S with density matrix \( \rho  \) we get \( \widetilde{\rho } \)
by a complete measurement, in the end, equivalent to the density matrix of a
type 2-system or an ensemble. 
\item Or vice versa, given \( \widetilde{\rho } \), - the density matrix belonging
to an ensemble - we reconstruct the state \( \rho  \) (\ref{dichte1}) of the
quantum object S from the diagonal elements of \( \widetilde{\rho } \). 
\end{enumerate}
Let us now define the off-diagonal information, the ``surplus-knowledge''
contained in the density matrix \( \rho  \) of quantum object S, as

\begin{equation}
\label{def_qsk}
K^{S}_{Q}:=Ctr(\rho -\widetilde{\rho })^{2}
\end{equation}

\( K_{Q}^{S} \) can be expressed, as desired, in terms of the off-diagonal
elements of the density matrix \( \rho  \):

\[
K_{Q}^{S}=Ctr(\rho -\widetilde{\rho })^{2}=C\sum _{i\neq j}\left| a_{i}a_{j}^{*}\right| ^{2}\]

resp.
\begin{equation}
\label{sec:def_qsk2}
K_{Q}^{S}=C\sum _{i}\left| a_{i}\right| ^{2}(1-\left| a_{i}\right| ^{2})=C-C\sum _{i}\left| a_{i}\right| ^{4}=C-\widetilde{I_{Q}}\quad \mbox {where}\quad \widetilde{I_{Q}}=Ctr\widetilde{\rho }^{2}
\end{equation}

This expression admits two interpretations:

\begin{enumerate}
\item \( K_{Q}^{S} \) may be interpreted as the difference between the information
obtained from the ontological and from the epistemical interpretation of a quantum
object, because \( \widetilde{I_{Q}}=Ctr\widetilde{\rho }^{2} \) reflects the
epistemological knowledge contained in the quantum object in question. Since
\( \widetilde{\rho } \) is diagonal we furthermore have \( Ctr(\rho -\widetilde{\rho })^{2}=Ctr\rho ^{2}-Ctr\widetilde{\rho }^{2} \).
\\
Hence, the whole information \( I_{Q} \) of a quantum object can be divided
into a classical part - contained in the diagonal elements - and a quantum part
- contained in the off-diagonal elements, i.e. \( I_{Q}=Ctr\rho ^{2}=C=K_{Q}^{S}+\widetilde{I_{Q}} \).
I again want to stress that the quantum part of the information depends on the
relation of state and measurement, i.e. the measured observable. There \emph{always}
is a measurement relative to which a quantum object is in a \emph{pure} state
(\ref{sec: unitary}). But simultaneously it is undetermined with respect to
non-commuting observables (s.a. the example in \ref{sec: beispiel}). In the
first case there is no surplus knowledge, \( K^{S}_{Q}=0 \), (relative to the
fixed measurement, which with probability 1 shows a fixed value for the measured
observable), but in the second case \( K^{S}_{Q}\neq 0 \). Hence the occurence
of a non-vanishing surplus-knowledge is deeply connected to the uncertainty
relations. 
\item In a second interpretation the surplus-knowledge \( K_{Q}^{S} \) can be regarded
as the possible information gain during a measurement or the information exchange
between the quantum object and its environment (resp. measuring apparatus):
If \( \rho  \) describes an (unknown) object before measurement and \( \widetilde{\rho } \)
the (partly) known object after a measurement then \( Ctr\rho ^{2} \) is the
information contained in the unmeasured quantum object (normally equal to \( C \))
and \( \widetilde{I_{Q}}=Ctr\widetilde{\rho }^{2} \) is the information still
contained in the quantum object after measurement. This would correspond to
building a partial trace in the standard density matrx formalism.
\end{enumerate}

\section{Working with the notion of information}

The sense and function of these notion can best be explored at work.

\subsection{Examples}

As the simplest possible example we treat the case of one resp. two photons.

\subsubsection{Case of single photon}

A single photon can be written as \( \psi _{1}=a_{1}\left| 0\right\rangle +a_{2}\left| 1\right\rangle  \)
with \( \left| a_{1}\right| ^{2}+\left| a_{2}\right| ^{2}=1 \). This corresponds
- relative to the standard basis \( \left| 0\right\rangle =\left( \begin{array}{c}
1\\
0
\end{array}\right) ;\, \left| 1\right\rangle =\left( \begin{array}{c}
0\\
1
\end{array}\right)  \) - to a density matrix 
\[
\rho _{1}=\left( \begin{array}{cc}
a_{1}a_{1}^{*} & a_{1}a_{2}^{*}\\
a_{1}^{*}a_{2} & a_{2}a_{2}^{*}
\end{array}\right) \]

The quantum information is \( I_{Q}=1 \) and the quantum mechanical surplus-knowledge
then is \( K_{Q}^{S}=2\left| a_{1}a_{2}\right| ^{2}\leq \frac{1}{2} \). It
is determined in relation to a spin measurement along the directions \( \left| 0\right\rangle ;\left| 1\right\rangle  \).
Relative to the representation of \( \psi  \) in the basis \( a_{1}\left| 0\right\rangle +a_{2}\left| 1\right\rangle , \)
\( a_{1}\left| 0\right\rangle -a_{2}\left| 1\right\rangle  \) the density matrix
would look like: \( \rho _{1}=\left( \begin{array}{cc}
1 & 0\\
0 & 0
\end{array}\right)  \) with no surplus knowledge, because the photon then is in a eigenstate relative
to the corresponding measurement. The relation between the surplus-knowledge
and the whole quantum information marks the amount of information extractable
from the quantum object - in a fixed context.

\subsubsection{Product state of two photons}

Let us assume that another photon \( \psi _{2}=b_{1}\left| 0\right\rangle +b_{2}\left| 1\right\rangle  \)
is brought into contact with the first photon. This can result in a type 1-system
which means that both photons get entangled and give rise to the most general
density matrix

\[
\rho =\rho _{1}\otimes \rho _{2}=\left( \begin{array}{cccc}
\left| a_{1}b_{1}\right| ^{2} & \left| a_{1}\right| ^{2}b_{1}b_{2}^{*} & a_{1}a_{2}^{*}\left| b_{1}\right| ^{2} & a_{1}a_{2}^{*}b_{1}b_{2}^{2}\\
\left| a_{1}\right| ^{2}b_{1}^{*}b_{2} & \left| a_{1}b_{2}\right| ^{2} & a_{1}a_{2}^{*}b_{1}^{*}b_{2} & a_{1}a_{2}^{*}\left| b_{2}\right| ^{2}\\
a_{1}^{*}a_{2}\left| b_{1}\right| ^{2} & a_{1}^{*}a_{2}b_{1}b_{2}^{*} & \left| a_{2}b_{1}\right| ^{2} & \left| a_{2}\right| ^{2}b_{1}b_{2}^{*}\\
a_{1}^{*}a_{2}b_{1}^{*}b_{2} & a_{1}^{*}a_{2}\left| b_{2}\right| ^{2} & \left| a_{2}\right| ^{2}b_{1}^{*}b_{2} & \left| a_{2}b_{2}\right| ^{2}
\end{array}\right) \]
 which already is properly normalized with quantum information \( I_{Q}=2 \)
and surplus knowledge

\( K_{Q}^{S}=4(\left| a_{1}\right| ^{4}\left| b_{1}b_{2}\right| ^{2}+\left| a_{1}a_{2}\right| ^{2}\left| b_{1}\right| ^{4}+2\left| a_{1}a_{2}b_{1}b_{2}\right| ^{2}+\left| a_{1}a_{2}\right| ^{2}\left| b_{2}\right| ^{4}+\left| a_{2}\right| ^{4}\left| b_{1}b_{2}\right| ^{2}) \)

\( =4((\left| a_{1}b_{1}\right| ^{2}+\left| a_{2}b_{2}\right| ^{2})(\left| a_{1}b_{2}\right| ^{2}+\left| a_{2}b_{1}\right| ^{2})+2\left| a_{1}a_{2}b_{1}b_{2}\right| ^{2})\leq \frac{3}{2} \).

If one of the photons would be in a eigenstate (e.g. \( b_{1}=1;b_{2}=0 \))
this surplus knowledge would reduce to \( 4\left| a_{1}a_{2}\right| ^{2}\leq 1 \). 

In general a system of \( n \) 2-state quantum objects with equal probabilities
\( \frac{1}{2} \) for all outcomes of a measurement possesses the (maximal
possible) surplus-knowledge 
\[
K_{Q,max}^{S}=n(1-\frac{1}{2^{n}})\]

\subsubsection{EPR-pairs of photons}

EPR-pairs are of special interest. Their density matrix can not be written in
terms of the product of the density matrices of the single potons (this is the
way they are constructed). Let us assume the singlett state \( \psi =\frac{1}{\sqrt{2}}(\left| 0,1\right\rangle -\left| 1,0\right\rangle ) \).
Herewith 
\[
\rho =\frac{1}{2}\left( \begin{array}{cccc}
0 & 0 & 0 & 0\\
0 & 1 & -1 & 0\\
0 & -1 & 1 & 0\\
0 & 0 & 0 & 0
\end{array}\right) \]
with respect to the basis built by eigenstates of the operator \( \sigma _{z} \).
This gives \( I_{Q}^{\mbox {EPR}}=2 \) and the \emph{maximal possible} surplus-knowledge
\( K_{Q}^{\mbox {EPR}}=1 \). One \emph{single} measurement can make the whole
system ``classical'' i.e. well determined with respect to this fixed measurement,
the \( \sigma _{z} \)- observable (see below section \ref{sec: criterion}).
\\
The same is valid for the so-called GHZ-states which have a similar density
matrix in the non-zero parts of their density matrix. Correspondingly \( I_{Q}^{\mbox {GHZ}}=3 \)
and \( K_{Q}^{\mbox {GHZ}}=\frac{3}{2} \). As seen below, (sections \ref{sec: criterion},
\ref{sec: assumption1}) also this system becomes ``classical'' with respect
to a fixed measurement observable in a single measurement.

\subsubsection{Ensemble of identical photons}

The density matrix of an ensemble of \( n \) identical photons is built as
the sum of the single density matrices. For further analysis I separately introduce
arbitrary phases \( \varphi _{i} \) such that
\[
\rho _{i,single}=\left( \begin{array}{cc}
a^{2}_{1} & a_{1}a_{2}e^{\imath \varphi _{i}}\\
a_{1}a_{2}e^{-\imath \varphi _{i}} & a_{2}^{2}
\end{array}\right) \]
Hence 
\[
\rho _{system}=\frac{1}{n}\sum _{i=1}^{n}\rho _{i,single}=\frac{1}{n}\left( \begin{array}{cc}
na_{1}^{2} & a_{1}a_{2}\sum _{i=1}^{n}e^{\imath \varphi _{i}}\\
a_{1}a_{2}\sum _{i=1}^{n}e^{-\imath \varphi _{i}} & na_{2}^{2}
\end{array}\right) \]
 The constant \( C \) is exactly \( 1 \) in this case. Hence there is the
surplus knowledge 
\begin{eqnarray*}
K_{Q}^{S}(\rho _{system})=\frac{1}{n^{2}}\left[ \left| \sum _{i}a_{1}a_{2}e^{\imath \varphi _{i}}\right| ^{2}+\left| \sum _{i}a_{1}a_{2}e^{-\imath \varphi _{i}}\right| ^{2}\right]  &  & \\
=\frac{\left| a_{1}a_{2}\right| ^{2}}{n^{2}}\left[ \left| \sum _{i}e^{\imath \varphi _{i}}\right| ^{2}+\left| \sum _{i}e^{-\imath \varphi _{i}}\right| ^{2}\right]  &  & 
\end{eqnarray*}

and the quantum information 
\[
I_{Q}=(a_{1}^{4}+a_{2}^{4})+\frac{a_{1}^{2}a_{2}^{2}}{n^{2}}\sum _{i,j=1}^{n}e^{\imath (\varphi _{i}-\varphi _{j})}=(a_{1}^{4}+a_{2}^{4})+\frac{2a_{1}^{2}a_{2}^{2}}{n^{2}}\sum _{i,j=1}^{n}\cos (\varphi _{i}-\varphi _{j})\]
 Now two extreme cases can be distinguished:

\begin{enumerate}
\item We assume \( n \) is very large and the phases \( \varphi _{i} \) are distributed
randomly with equal weight. Then the term containing the diagonal elements outweighs
the other term depending on the phases. Hence \( I_{Q}\simeq (a_{1}^{4}+a_{2}^{4})<1 \)
and \( K_{Q}^{S}\simeq 0 \). This indicates a transition from a quantum system
to a ``classical'' system where the non-knowing of measurement results can
be interpreted epistemically.
\item We assume all the phases \( \varphi _{i} \) are equal to a single phase \( \varphi  \)
. Then the expressions for the informations simplify to \( K_{Q}^{S}=2\left| a_{1}a_{2}\right| ^{2} \),
the surplus knowledge contained in a single photon, and to \( I_{Q}=(a_{1}^{2}+a_{2}^{2})^{2}=1 \).
Taken together this indicates that the ensemble constitutes a quantum object
with only two possible states, corresponding to an ensemble of coherent photons,
behaving like \emph{one} single photon. In this place it is quite interesting
to note that hence an ensemble of identical photons as required e.g. in the
ensemble interpretation, cannot be distinguished from a single photon. Both
carry the same information content and the same surplus-knowledge.
\end{enumerate}
For convenience I give the formulas for a system of two identical photons which
already display all the described behaviour: the density matrix is 
\[
\rho _{system}=\frac{1}{2}\left( \begin{array}{cc}
2a_{1}^{2} & a_{1}a_{2}(1+e^{\imath \varphi })\\
a_{1}a_{2}(1+e^{-\imath \varphi }) & 2a_{2}^{2}
\end{array}\right) \]
 with \( K_{Q}^{S}=\frac{\left| a_{1}a_{2}\right| }{2}^{2}(1+\cos \varphi ) \)
and \( I_{Q}=(a_{1}^{4}+a_{2}^{4})+\frac{\left| a_{1}a_{2}\right| ^{2}}{2}(\cos \varphi +1) \),
where \( \varphi  \) is the relative phase of the two photons. If \( \varphi =0 \),
then the density matrix of the whole system \( \rho _{\mbox {system}} \) is
just that of the single photon's density matrix \( \rho _{\mbox {single}} \).
This again shows that two identical photons together are described by \emph{one}
wavefunction, exhibiting single particle behaviour. A similar phenomenon occurs
in the Bose-Einstein-condensates.

\subsection{Quantum Information and Measurement }

With aid of these notions of quantum information and surplus knowledge we now
approach the measurement problem. We do not go into any detailed discussion
of the measurement problem; this can be read elsewhere \cite{Messproblem}.

\subsubsection{Assumptions for measurement}

By definition we cannot know anything about pure quantum objects, i.e. objects
of type 1. Hence we make

\begin{description}
\item [Assumption1:]Only quantum objects of type 2 can be used as a measuring apparatus
(see also \cite{Messproblem}). 
\end{description}
By the very definition we can have knowledge only about quantum objects of type
2, because they allow for an epistemical interpretation of quantum objects;
i.e. there are fixed values for the properties to be measured, the observer
only does not know which value is realized. Objects of type 1 in contrast do
not have fixed values for its properties at all; the properties of those objects
come into existence only with a measurement. 

Hence the measurement problem is most deeply related to interaction case 3 from
section \ref{sec: interaction}. 

Now we assume quantum objects with density matrices as in \ref{dichte1} and
in \ref{dichte_ensemble} and using the same notation we state the reduction
postulate as\label{sec: assumption1}:

\begin{description}
\item [Assumption2:]If after the measurement any \( a_{j}^{M}=0 \) (that has been
different from 0 before measurement), then there are at least one index \( i_{1},....,i_{r} \)
such that \( a_{i_{1}}^{S}=...=a^{S}_{i_{r}}=0 \).
\item [Remark1:]This assumption goes just the other way round than most other assumptions
on measurement devices. In my opinion this formulation gives the possibility
of dealing with the phenomenon of so-called quantum erasers. 
\item [Remark2:]A measurement is only fruitful and hands over new information from
the quantum object to the classical regime if more than one of the \( a_{i}^{M} \)or
\( a_{i}^{S} \) are different from zero. A measuring apparatus \( M \) should
give a statement which allows to draw conclusions on the quantum object \( S \).
I.e. if a possible result of \( M \) is excluded with probability one (that
is one \( a_{j}^{M}=0 \)), then there should be properties of \( S \) that
also can be excluded with probability one. This seems to me to be a reasonable
assumption because otherwise any measurement would be completely useless or
put differently: the result \( a_{j}^{M}=0 \) of \( M \) would give no information
on \( S \), i.e. it would be no true measurement.
\end{description}
As we have seen, the information that can be extracted from a quantum object
depends on the design of the measurement or - more generally - on the environment
it is brought into contact with. Furthermore - in order to extract and interpret
the information - we have to know something about the measuring device.

\subsubsection{Criterion for completion of measurement}

Given a state and a fixed measurement (observable) the quantum mechanical surplus-knowledge
\( K_{Q}^{S} \) measures the quantum object's degree of being ``quantum''
with respect to this measurement. From the definition of \( K_{Q}^{S} \) we
can define a quantum object as ``behaving classical'' if \( K_{Q}^{S} \)
is sufficiently small. Classically there cannot be an amount of information
less than 1 bit; so we set:

\begin{description}
\item [Criterion:]The object S can be regarded as a classical object with respect
to a fixed measurement if the corresponding surplus knowledge \( K_{Q}^{S}<1. \)
Then we regard the measurement as completed (at least as far as possible)\label{sec: criterion}.
\item [Remark:]This condition is always fulfilled in case of a 2-dimensional Hilbert
space, i.e. e.g. for the spin states of a single photon. But already for 3-dimensional
Hilbert space, there has to be some measurement in order to reduce the quantum
mechanical surplus knowledge before a quantum object can be regarded as ``classical'',
i.e. as having fixed values for one property.
\end{description}

\subsection{Example for Measurement: Photon Interference}

Now let us treat the most important case, the case of a photon being measured
by a object showing two possibilities. This setting is given for example by
the double slit experiment, a Michelson interferometer or similar devices. 

Of special interest in this context is the treatment of the phenomenon of quantum
eraser and which-way-information (\cite{zeilinger1}).

As alluded to before an interference experiment would correspond to case 3:
the interaction of the photon with a measuring device having two well defined
possibilities (ways).

\paragraph{Before the measurement: }

The density matrix of an interferometer (before the measurement) is given by:
\[
\rho _{M}=\left( \begin{array}{cc}
\alpha ^{2} & 0\\
0 & \beta ^{2}
\end{array}\right) \quad \mbox {with}\quad \alpha ^{2}+\beta ^{2}=1\]
 and the density matrix of the photon is as described above (see (\ref{dichte1})).
Then the common density matrix of photon and measuring device is given by 
\[
\rho _{S}\otimes \rho _{M}=\left( \begin{array}{cccc}
\alpha ^{2}a_{1}a_{1}^{*} & \alpha ^{2}a_{1}a_{2}^{*} & 0 & 0\\
\alpha ^{2}a_{1}^{*}a_{2} & \alpha ^{2}a_{2}a_{2}^{*} & 0 & 0\\
0 & 0 & \beta ^{2}a_{1}a_{1}^{*} & \beta ^{2}a_{1}a_{2}^{*}\\
0 & 0 & \beta ^{2}a_{1}^{*}a_{2} & \beta ^{2}a_{2}a_{2}^{*}
\end{array}\right) \]
 The definition of quantum information gives:

\[
I_{Q}^{I}:=2tr(\rho _{S}\otimes \rho _{M})^{2}=2I_{Q}^{M}=2(\alpha ^{4}+\beta ^{4})\]
 We distinguish two cases: 

\begin{enumerate}
\item \( \alpha ,\beta \neq 0 \). In this case \( I_{Q}^{I}<2 \), where the strict
inequality reflects the fact that the compound system is not a whole quantum
object but includes a ``semiclassical'' device. The smaller \( I_{Q}^{I} \)
the more the measurement device makes the quantum object ``classical''. On
the other hand the quantum object can not be interpreted completely as classical
as is seen from the surplus-knowledge \( K^{I}_{Q}:=2\sum _{i\neq j}\mbox {off-diagonal}^{2}=4(\alpha ^{4}+\beta ^{4})\left| a_{1}a_{2}\right| ^{2}=I_{Q}^{I}\cdot K_{Q}^{S} \),
indicating the quantum character of the compound system. The minimum of both,
\( I_{Q}^{I} \) and \( K_{Q}^{I} \), therefore is attained if both ``ways''
can be discerned clearly, i.e. \( \alpha ^{2}=\beta ^{2}=\frac{1}{2} \). Then
\( K_{Q}^{I}=2\left| a_{1}a_{2}\right| ^{2}=K_{Q}^{S} \), the surplus knowledge
of the single photon which only vanishes if \( a_{1}=0 \) or \( a_{2}=0 \)
meaning that the quantum object would be in a eigenstate with respect to the
chosen measurement. 
\item \( \alpha =1;\beta =0 \) or vice versa. In this case the measurement can give
no information on the quantum object to the outside environment and the quantum
information remains undisturbed. Here the surplus knowledge \( K_{Q}^{I} \)
becomes maximal - \( K_{Q}^{I}=4\left| a_{1}a_{2}\right| ^{2} \) - corresponding
to the fact, that - no matter what is done during the ``measurement'' - we
cannot distinguish different states of the quantum object in question. That
means there is no true measurement, because no information is extracted from
the quantum object.
\end{enumerate}
These information values characterize the compound system only before a measurement:
the quantum object interacts with the measurement device without being read
from the outside.

\paragraph{After the measurement: }

After the measurement of a single photon, we should know ``which way it has
taken'', simulating the situation as if (for this single photon) \( \alpha =1 \)resp.
\( \beta =0 \) and from this information we would like to draw conclusions
on the quantum object. A measurement on an ensemble requires that for every
single photon it has to be decided whether ``the photon has taken way 1''
or whether ``the photon has taken way 2'', giving in the end the respective
probabilities \( \alpha ^{2} \) resp. \( \beta ^{2} \). Hence from (the factual)
\( \beta  \) equal to 0 for a single photon we should be able to conclude e.g.
(the fact) \( a_{1}=0 \), according to assumption 2 (reduction postulate, \ref{sec: assumption1}).
Then \( I_{Q}^{I,\mbox {after}}=2 \), that is: M carries the whole information
of the interaction and \( K^{S,\mbox {after}}_{Q}=4\left| a_{1}a_{2}\right| ^{2}=0 \).
Then the measurement is completed and the photon may be regarded as a classical
object with definite properties \emph{- but!: with respect to the performed
measurement only! }

But since \( I_{Q}^{I,\mbox {after}} \) has attained its maximal possible value,
there might still be non-vanishing surplus knowledge with respect to other measurements
(observables).

If, as for instance in a quantum eraser, assumption 2 (see \ref{sec: assumption1})
is hurt, i.e. \( \beta =0 \) \emph{and} \( a_{1}\neq 0 \) after a measurement,
\( K^{S,\mbox {after}}_{Q} \) is different from 0, i.e. the photon is \emph{not}
in a eigenstate with respect to the basis of the measurement or, in other words,
(part of) the information is left to the quantum object.

\section{Conclusion}

In this article there were introduced two notions of quantum information reflecting
the differences between quantum and classical objects. These give the fundamental
notion of information a quantitative expression and show clearly the contextuality
of quantum objects.

\begin{description}
\item [Acknowledgement:]I have to thank Prof. Thomas Görnitz for innumerable helpful
discussions and many valuable hints.
\end{description}

\bibliographystyle{alpha}

\bibliography{lit}

\end{document}